# Properties of metal-insulator transition and electron spin relaxation in GaN:Si

A. Wolos

*Institute of Physics Polish Academy of Sciences, Al. Lotnikow 32/46, 02-668 Warsaw, Poland*

Z. Wilamowski

*Institute of Physics Polish Academy of Sciences, Al. Lotnikow 32/46, 02-668 Warsaw, Poland*

*Faculty of Mathematics and Computer Sciences, University of Warmia and Mazury, ul. Zolnierska 14, 10-561 Olsztyn, Poland*

M. Piersa and W. Strupinski

*Institute of Electronic Materials Technology, ul. Wolczynska 133, 01-919 Warsaw, Poland*

B. Lucznik, I. Grzegory, and S. Porowski

*Institute of High Pressure Physics of the Polish Academy of Sciences, Unipress, ul. Sokolowska 29/37, 01-142 Warsaw, Poland*

ABSTRACT:

We investigate properties of doping-induced metal-insulator transition in GaN:Si by means of electron spin resonance and Hall effect. While increasing the doping concentration, Si-related bands are formed below the bottom of the GaN conduction band. The $D^0$ band of single-occupied Si donor sites is centered 27 meV below the bottom of the GaN conduction band, the $D^-$ band of double-occupied Si states at 2.7 meV below the bottom of the GaN conduction band. Strong damping of the magnetic moment occurs due to filling of the $D^-$ states at Si concentrations approaching the metal-insulator transition. Simultaneously, shortening of electron spin relaxation time due to limited electron lifetime in the single-occupied $D^0$ band is observed. The metal-insulator transition occurs at the critical concentration of uncompensated donors equal to about $1.6 \times 10^{18}$ cm$^{-3}$. Electronic states in metallic samples beyond the metal-insulator transition demonstrate non-magnetic character of double-occupied states.

## 1. INTRODUCTION

Gallium nitride is a wide band gap semiconductor finding application in blue and UV optoelectronic,[1] high-power electronic,[2] and due to weak spin-orbit interaction and expected long electron spin coherence it is also investigated for a use in spintronics.[3,4] There is also considerable interest in doping GaN with transition metals, in order to combine magnetic and semiconducting properties, likewise for use in spintronic devices.[5] Silicon as a popular shallow donor in GaN is serving as a source of effective mass electrons, allowing obtaining n-type material. As both electron transport and spin coherence in n-type semiconductors are practically determined by properties of donor bands, we present systematic studies of the Si doping-induced bands in GaN. The investigations were performed by means of electron spin resonance (ESR) and Hall effect. In particular, we focus on properties of GaN:Si being close to the metal-insulator transition (MIT).

The critical Si concentration for the MIT in GaN has been already theoretically estimated,[6,7] however the exact nature of the transition has not been investigated experimentally. In this paper we present a classical analysis of electron transport close to the MIT, evaluating activation energies of the resistivity and the Hall concentration. Their dependence on the doping allows us to determine critical concentration for the MIT. The electron transport data are analyzed within Shklovskii's formalizm,[8] which distinguishes the three activation energies: $E_1$-to the GaN conduction band, $E_2$-the activation energy to double-occupied donor states, and $E_3$-the hopping activation energy. Such analysis is presented in Chapter 5.

Electron spin resonance provides deeper understanding of the nature of the MIT in GaN:Si. The measurements unambiguously show that electronic magnetic moment is damped when approaching the critical concentration for the MIT, which is consistent with filling non-magnetic double-occupied donor states ($D^-$). Simultaneously, electron lifetime in the single-occupied donor band ($D^0$) becomes shortened due to excitations to the $D^-$ band. This influences also electron spin relaxation processes. In metallic samples being well beyond the MIT all the conducting electrons occupy non-magnetic electronic states.

## 2. SAMPLES AND EXPERIMENTAL DETAILS

Two sets of GaN:Si crystals were studied. Epilayers were grown by Metal-Organic Chemical Vapor Deposition (MOCVD) on a sapphire substrate, with donor concentration varying between $10^{17}$ $cm^{-3}$ and $10^{19}$ $cm^{-3}$. Epilayers were thinner than 2 microns. Bulk samples were obtained by Hydride Vapor Phase Epitaxy (HVPE) on high pressure-grown plate-like GaN seeds. Details of the growth procedure are summarized in Ref. 9. Thickness of the HVPE samples ranged up to 1 mm, while lateral dimensions were typically $4 \times 4$ $mm^2$. Dislocation density ranged up to $10^7$ $cm^{-3}$.[10] Si concentration varied between $10^{17}$ $cm^{-3}$ and $10^{19}$ $cm^{-3}$. Large volume of the HVPE-crystals allowed us to perform precise ESR measurements.

ESR measurements were performed using Bruker ESP 300 spectrometer operating in X-band with microwave frequency of 9.5 GHz. Temperature was varied down to 2.5 K using an Oxford continuous-flow cryostat. Magnetic field was calibrated with a DPPH marker.

## 3. ELECTRON SPIN RESONANCE OF EFFECTIVE MASS DONORS IN GaN

ESR of residual donors in wurtzite GaN films grown by MOCVD has been studied by Carlos and coworkers.[11] The anisotropic g-factor of effective mass donors ($D^0$) has been determined to be equal to $g_{\|}$ = 1.951 and $g_{\perp}$= 1.948, while the mean g-value has been explained using 5-band k·p model. Resonance lines characterized by this g-factor have been next reported in crystals obtained by the different techniques: MOCVD-, MBE-, and HVPE-grown GaN doped with Si, or intentionally undoped, as well as in intentionally undoped AMMONO-GaN micropowders.[12,13] In this work we extended the ESR studies of GaN on highly Si-doped samples grown by HVPE to study properties of the metal-insulator transition.

ESR spectra of Si donors in HVPE-GaN (T = 2.5 K) are shown in Fig. 1. The resonance recorded for a sample with the lowest silicon concentration in a series, $n_{Si}$-$n_a$ = $3.3 \times 10^{16}$ $cm^{-3}$ (where $n_{Si}$ denotes total Si concentration, $n_a$ concentration of compensating acceptors, the parameters will be determined from Hall measurements in subsequent sections), has a symmetric Lorentzian lineshape. For higher Si concentrations, $n_{Si}$-$n_a$ = $1.5 \times 10^{18}$ and $1.6 \times 10^{18}$ $cm^{-3}$, the lineshape is Dysonian, characteristic for conducting samples. For a sample with the highest Si concentration, $n_{Si}$-$n_a$ = $5.3 \times 10^{18}$ $cm^{-3}$, we did not record any resonance.

The resonance lineshapes become modified while elevating temperature, reflecting changes in GaN:Si conductivity. We performed an analysis of the ESR signals exploiting the Dyson's formalism.[14,15] This allowed us to determine skin depth $\delta$ for microwave penetration, which at T=2.5 K is equal to > 100 mm, 0.2 mm, and 0.14 mm for samples with $n_{Si}$-$n_a$ equal to 3.3 × $10^{16}$ $cm^{-3}$, 1.5 × $10^{18}$, and 1.6 ×$10^{18}$ $cm^{-3}$, respectively. The performed analysis will allow us to subtract contribution of the skin effect from the amplitude of the ESR signals discussed in Chapter 6.

## 4. ELECTRON SPIN RELAXATION IN GaN:Si

Electron spin relaxation in n-type GaN has been recently investigated by time-resolved optical methods.[3,4] The authors of Ref. 3 have observed the non-monotonic variation of spin lifetimes with the doping, with the longest low-temperature spin lifetime equal to 20 ns for intermediate doped samples. Interestingly, the measured spin coherence appeared to be robust to the presence of large number of dislocations.

Electron spin resonance provides classical methods for determination of transverse ($T_2$) and longitudinal ($T_1$) spin relaxation times. A resonance linewidth is connected to spin relaxation through a relation: $\Delta B_{HWHM} = 1/(\gamma T_2) = 1/(2\gamma T_1) + 1/(\gamma T_2^{'})$, where $\gamma$ is electron gyromagnetic ratio and $T_2^{'}$ denotes a dephasing time related to all effects other than these for the $T_1$.

The longitudinal spin relaxation time for a sample with $n_{Si}$-$n_a$ = 3.3 × $10^{16}$ $cm^{-3}$ was estimated from the ESR signal using a standard saturation method.[16,17] Figure 2(a)-(b) shows the dependence of the integrated amplitude and of the linewidth on microwave power. At low microwave powers the amplitude grows like square root of the power. When the power absorbed is comparable to the power relaxed by the system the amplitude saturates. Saturation effects are visible also as the broadening of the resonance line (Fig. 2(b)). The saturation method has been described in classical ESR textbooks, where one can find formulas to fit the microwave power dependence, see e.g. Refs. 16 and 17. Fitting the power dependence of the data shown in Fig. 2 one gets $T_1$ equal to about 3000 ns at the temperature of 2.5 K, and 12 000 ns at the temperature of 40 K. This gives the contribution to the broadening of the resonance line less than 0.01 G in both cases. The measured resonance linewidth is higher than 3 G in a whole temperature range, thus the contribution of longitudinal relaxation to the linewidth can be neglected. For samples with higher Si concentration the amplitude of the resonance does not saturate up to the microwave power of 200 mW. This means the $T_1$ to be shorter than about 3 000 ns. The contribution of the $T_1$ to the linewidth cannot be in these cases excluded.

The dependence of the ESR linewidths on temperature for a set of HVPE-GaN:Si is shown in Fig. 3. The sample with the lowest Si concentration ($n_{Si}$-$n_a$ = 3.3 × $10^{16}$ $cm^{-3}$) is very typical.[11] The resonance line, which is inhomogenous broadened due to fluctuations of effective magnetic field felt by the donor electrons (e.g. originating from super-hyperfine interactions with spins of neighboring nuclei), undergoes motional narrowing while elevating temperature. The linewidth narrows from 10 G ($T_2$ elongates from 5.6 ns) at T = 2.5 K down to 2.7 G (21 ns) at T = 30 K. The narrowing mechanism is related to thermally activated motion of electrons leading to averaging of effective magnetic field fluctuations, thus to narrowing of the resonance. Typically, activation energies of the motional narrowing are equal to the activation energies for the resistivity. This is clearly visible, *e.g.* on an example of ESR in n-type ZnO.[18] In a case of GaN the effect of motional narrowing must be shadowed by other relaxation mechanisms. The activation energy of the narrowing determined for $n_{Si}$-$n_a$ = 3.3 × $10^{16}$ $cm^{-3}$ sample (0.5 meV) is too low to be attributed solely to the motional effect. The activation energy for the resistivity is expected to be of the order of a few meV in this case (see Chapter 5).

The ESR linewidth undergoes subsequent broadening with activation energy of 11 meV while further elevating temperature. The $T_2$ shortens down to 4.4 ns at T = 50 K. The broadening is correlated with excitation of the donor electrons to the GaN conduction band, promoting interaction with GaN lattice (acoustic) phonons.[11]

While increasing donor doping (reducing distance between the donor sites) the magnitude of exchange interactions is increased. This leads to averaging of the local magnetic field for the ensemble of exchange coupled electrons. As a result the inhomogeneous contribution to the linewidth is eliminated, narrowing the resonance line.[19] Indeed, this effect is observed in a sample with $n_{Si}$-$n_a$ = 1.5 × $10^{18}$ $cm^{-3}$, for which the low-temperature linewidth is reduced down to about 4.3 G ($T_2$ = 13.6 ns), and no further narrowing effects are visible with varied temperature.

The two samples with higher Si doping are qualitatively different from lighter doped crystals. Besides homogeneous character of the resonance line, a new broadening effect appears. A shortening of the electron spin relaxation time $T_2$ is clearly visible both with increasing the Si doping and with elevating temperature. The low-temperature $T_2$ shortens from 13.6 ns for the $n_{Si}$-$n_a$ = 1.5 × $10^{18}$ $cm^{-3}$ sample down to 6.8 ns for the $n_{Si}$-$n_a$ = 1.6 × $10^{18}$ $cm^{-3}$ sample. The effect of temperature is even more pronounced, reducing the spin relaxation time to 1.7 ns at T = 20 K for the latter sample.

The ESR results are in good agreement with these obtained in Ref. 3 by time resolved Faraday rotation, despite the fact that optical methods may be influenced by processes of carrier excitation and recombination. The longest spin relaxation times $T_2$ in our samples are observed either due to exchange or due to motional narrowing, which mechanisms remove effects of inhomogeneous broadening. The resulting low-temperature spin relaxation times are of the order of 14-21 ns, similar as reported in Ref. 3. One can attempt to assign the long spin relaxation times to the fundamental spin-orbit interactions in GaN. The interplay between (i) broadening due to super-hyperfine interaction with neighboring nuclei, (ii) exchange narrowing, and (iii) motional narrowing leads to the non-monotonic variation of spin relaxation times with the doping. Additionally, (iv) a new mechanism becomes pronounced when approaching metal-insulator transition. A drastic shortening of the $T_2$ has been observed in both time-resolved Faraday rotation and in the ESR experiment. We will relate to this problem in Chapter 6.

## 5. HALL MEASUREMENTS

To illustrate formation of Si donor bands in GaN we performed Hall measurements on both HVPE-grown GaN:Si and MOCVD-grown epilayers. Symbols in Fig. 4 show Hall concentration, mobility and the resistivity measured versus temperature. The results are very typical for impurity-related conduction in a standard semiconductor. One can describe the Hall data by a classical Shklovskii's formalism based on early works of Mott and Hubbard.[8] Solid lines in Fig. 4 are fitted according to this model, employing a fitting procedure the same as in Ref. 20. Table I collects the obtained Hall parameters, which are: concentration of uncompensated Si donors, $n_{Si}$-$n_a$, compensation ratio, $K = n_{Si}/n_a$, and three activation energies, $E_1$, $E_2$ and $E_3$.

Electric conductivity is decomposed here to the three mechanisms: conduction in the host crystal conduction band activated with energy $E_1$, and two mechanisms related to the conduction due to donor impurities. A hopping conduction relates to tunneling of electrons between occupied ($D^0$) and empty ($D^+$) donor sites. It is thermally activated with energy $E_3$, and characterized by a negligent effective Hall mobility. The third conduction mechanism distinguished here is related to the motion of electrons over singly-filled neutral donor sites, with activation energy denoted by most authors by $E_2$. The extra electron occupying neutral donor site forms a $D^-$ center. Because of a large radius of $D^-$ centers, they overlap strongly at intermediate impurity concentration, allowing electric transport. Simultaneously, the mobility in the $D^-$ band is considerably higher than for the hopping conduction. The $D^-$ band of doubly-occupied donors has been sometimes identified with the upper Hubbard subband.[8,21, 22,23]

Basing on the above model the following features of electron transport in GaN:Si are recognized. The high-temperature conductivity is dominated by electron conduction in the GaN conduction band, activated with energy $E_1$. The $E_1$, which for compensated lightly doped samples corresponds to the ionization energy of an isolated donor, $E_0$, decreases systematically while increasing Si doping due to broadening of the $D^0$ band. Shklovskii gives a following formula for the $E_1$:[8]

$$E_1 = E_0 - f(K)\frac{e^2}{4\pi\ \varepsilon\varepsilon_0} n_{donor}^{1/3}. \qquad \text{Eq. 2.}$$

Here, the activation energy depends on the electrostatic energy of electrons separated by a distance equal to the average distance between donor sites. Effects of compensation are taken into account by including f(K)-a function of the compensation parameter K. For low silicon concentration limit the $E_1$ tends to $E_0 = 27$ meV in GaN, which is in good agreement with the optical Si level, 22-29 meV.[24,25] $E_1$ takes value of about 21.5 meV for HVPE-sample with Si concentration $n_{Si}$-$n_a$ = 3.3 × $10^{16}$ $cm^{-3}$, and tends to zero at the critical concentration of about $n_{Si}$-$n_a$ = 1.6 × $10^{18}$ $cm^{-3}$ (Fig 5).

At low temperatures, a classical nearest-neighbor hopping is observed in lightly doped samples having $n_{Si}$-$n_a$ below 1.0 × $10^{17}$ $cm^{-3}$. Hopping appears in Hall measurements as rapid drop of effective mobility and simultaneous apparent increase in Hall concentration. This occurs for $n_{Si}$-$n_a$ = 1.0 × $10^{17}$ sample at the temperature of about 40 K. Below this temperature hopping conduction dominates, mobility is too low to be measured and effective Hall concentration returns random values. The thermal activation energy $E_3$ equals to 4.5 meV for this sample. Such behavior is not observed in stronger doped samples.

In MOCVD-grown samples with $n_{Si}$-$n_a$ = 6.0 × $10^{17}$ $cm^{-3}$ and 1.6 × $10^{18}$ $cm^{-3}$ there is lack of the typical hopping transport at low temperatures. Instead, the conductivity shows low but definite mobility, allowing attributing the dominating conductivity mechanism to the $E_2$-type of electron conduction. The $E_2$ mechanism has been often attributed to the motion of electrons over single-filled neutral donors, so in the $D^-$ band.[8,22] The $D^-$ notation originates from the physics of defects in solids and denotes a negatively charged donor in a substitutional position.[21] In Mott-Hubbard model of metal-insulator transition it is considered that an extra electron can be added to the lattice of single-filled donor sites, resulting in splitting of the donor band into two subbands: lower Hubbard subband of single-filled states and upper Hubbard subband of double-filled states.[23] The subbands are separated by the energy U (the Hubbard U) equal to the interaction energy of two electrons of opposite spins located at the same site. It is natural to link the Mott-Hubbard picture with defect models, to identify the upper Hubbard subband with the D- band of double-occupied donors. This relation has been often pointed out in literature.[8,22,26] From the Hall data one can determine location of the $D^-$ band in GaN:Si, which is about 2.7 meV below the bottom of the GaN conduction band (for $n_{Si}$-$n_a$ = 6.0 × $10^{17}$ $cm^{-3}$ sample). The $D^0$ band is placed about 27 meV below the bottom of the GaN conduction band, which gives splitting of the two silicon donor subbands equal to about 24.3 meV, which equals to 0.9 of the ionization energy of an isolated Si donor. This is a value fairly expected in the Mott-Hubbard description (U = 0.96 $E_0$ in the original model).[8, 6]

The activation energy $E_2$ in GaN:Si samples decreases with increasing Si doping. It is worth noting, that it tends to zero at the same critical concentration as vanishing of the $E_1$ activation energy, Fig. 5. The situation when the $E_1$ and the $E_2$ tend to zero at the same critical impurity concentration is not a universal rule for III-V semiconductors. It has been, *e.g.* earlier shown, that in GaAs doped with Mn acceptor the $E_2$ vanishes at the critical concentration of about 3 × $10^{19}$ $cm^{-3}$, above which metallic conduction occurs in the Mn acceptor band, while the $E_1$ tends to zero at much higher Mn concentration equal to about 2 × $10^{20}$ $cm^{-3}$.[20] Similar situation appears in antimony-doped germanium.[22] In our case, however, we cannot distinguish unambiguously two separate critical concentrations for vanishing of the $E_1$ and of the $E_2$, respectively. Although a slight slope in Hall concentration data is visible at high temperatures for all samples having concentration higher than $n_{Si}$-$n_a$ = 1.6 × $10^{18}$ $cm^{-3}$ (higher than the MIT) we cannot attribute this slope to the non-zero $E_1$ activation energy. It is related rather to the temperature dependence of the density of states. In non-degenerate semiconductors the concentration n of electrons in the conduction band can be approximately expressed as $n \sim N_c Exp(E_1/kT)$, where $N_c$ is the effective density of states in the conduction band dependant on temperature like $T^{3/2}$. One can see, that even for $E_1$ equal to zero the electron concentration in the conduction band is not independent on temperature. On the other hand for highly degenerated semiconductors one can show that the $T^{3/2}$ dependence cancels out and n becomes a constant number.[27] There exists a region of donor concentrations in the vicinity of the metal-insulator transition which represents intermediate properties between non-degenerate and highly-degenerate picture. Summarizing, it seems for us that both the activation energies $E_1$ and $E_2$ vanish at the same critical concentration.

It is also worth noting, that the critical concentration for the MIT depends on the compensation. We have accepted a value of $1.6 \times 10^{18}$ cm$^{-3}$, which is evidently an onset of metallic conduction in HVPE-grown samples. In MOCVD-crystals the sample with $n_{Si}$-$n_a$ = $1.6 \times 10^{18}$ cm$^{-3}$ still shows small thermal activation in conductivity, so the MIT occurs at somewhat higher Si concentration. This small discrepancy may be attributed to experimental error in determination of the Si concentration, or to the dependence on the compensation. Expression of the critical concentration in terms of a number of neutral donors, $n_{Si}$-$n_a$, allows us to take into account compensation effects, however may not be sufficient. It has been pointed out in Ref. 22 that the effect of compensation on $E_2$ involves more than an increase in the average separation between neutral donors.

Figure 6 shows schematic representation of the density of states introduced by the Si doping in GaN, shown on an example of a sample having $n_{Si}$-$n_a$ = $6.0 \times 10^{17}$ cm$^{-3}$. Both the $D^0$ band of single-occupied donor states and the $D^-$ subband of double-occupied states are shown. In order to explain the definite $E_2$ activation energy, the $D^-$ band is plotted as a narrow peak in the density of states, narrower than the $D^0$ subband. This is in contrast to the larger localization radius of the $D^-$ states, which should lead to a larger bandwidth. We will leave this inconsistency, as so far there is lack of a plausible explanation (see discussion in Ref. 8).

The critical Si concentration discriminating between metallic and thermally activated conductivity equals to about $1.6 \times 10^{18}$ cm$^{-3}$, which is in good agreement with theoretical considerations in Refs. 6 and 7. The electron transport measurements show, that the metal-insulator transition in GaN:Si has many features characteristic for the Mott-Hubbard type of the transition. Si donor band splits into two subbands $D^0$ of single- and $D^-$ of double-occupied sites. The MIT corresponds to closing the gap between the two subbands, simultaneously the activation energy to the GaN conduction band disappears. Beyond the MIT the two donor subbands and the GaN conduction band overlap and lose their structure. In the next chapter we will show that the two electrons at a double-occupied donor site are antiferromagnetic coupled. This will be concluded from ESR.

## 6. DOUBLE-OCCUPIED STATES - CONCLUSIONS FROM ESR

Figure 7 shows temperature dependence of the ESR amplitude in HVPE-grown GaN:Si. To obtain the amplitude of the signal the Dysonian lineshape was decomposed into absorption and dispersion part. The amplitude of the absorption part was plotted then in Fig. 7. Corrections for the skin effect were taken into account where necessary. The resonance amplitudes show at low temperatures a 1/T dependence characteristic for the Curie paramagnetism. The single-occupied Si donor states form a classical paramagnetic band. The concentration of Si donors being in a neutral $D^0$ state can be calculated from Fig. 7 by comparing the amplitude of the resonance to the calibrated standard sample. The concentration of $D^0$ centers determined in this way is plotted in Fig. 8 versus the $n_{Si}$-$n_a$ parameter.

The amplitude of the ESR is proportional to the number of paramagnetic centers in a sample. Neutral donor concentrations obtained in this way for n-type GaN reported in literature agree reasonably with the nominal doping.[11,13] This simple rule, however fails when approaching metal-insulator transition, Fig. 8. Indeed the concentration of neutral donors is proportional to $n_{Si}$-$n_a$ for the lightest doped sample ($3.3 \times 10^{16}$ $cm^{-3}$), but for the two samples close the metal-insulator transition ($1.5 \times 10^{18}$ $cm^{-3}$ and $1.6 \times 10^{18}$ $cm^{-3}$) the resonance amplitude is drastically reduced with respect to the $n_{Si}$-$n_a$. Finally, for the highest doped sample ($5.3 \times 10^{18}$ $cm^{-3}$) we did not record any ESR signal. The rapid drop of the ESR amplitude when approaching metal-insulator transition can be well explained by filling double-occupied $D^-$ states, occurring at the expense of $D^0$ centers. When spins of the two electrons occupying $D^-$ site are anti-parallel coupled, they form a non-magnetic center, not responsive to ESR technique. Solid line in Fig. 8 shows a following dependence:

$$n_{D^0} = (n_{Si} - n_a) \cdot f^{(1)}, \qquad \text{Eq.2.}$$

where $n_{D^0}$ is concentration of neutral Si donors (which is seen by ESR), $n_{Si}$-$n_a$ is a difference between total concentration of Si and concentration of acceptors, finally $f^{(1)}$ is the Gibbs distribution describing probability of finding one electron on the two-electron center. Parameters necessary to calculate the probability: position of the Fermi level and location of $D^0$ and $D^-$ levels are obtained from the Hall data. When the Fermi level is located between $D^0$ and $D^-$ levels the lower level ($D^0$) is full occupied and the upper level ($D^-$) is practically empty. The situation drastically changes when the Fermi level approaches the upper ($D^-$) level. The mass occupation of $D^-$ level occurs, which is followed by emptying of the $D^0$ level. Although the above description is for the energy levels, not bands, it can give simple quantitative understanding of diminishing the number of donors in neutral charge state occurring at the metal-insulator transition. It is worth to stress, that in metallic GaN:Si samples being beyond metal-insulator transition the electronic states formed from Si donor bands and the GaN conduction band are double-occupied with vanishing magnetic moment.

Excitation of the donor electrons from single-occupied $D^0$ band to doubly occupied $D^-$ band influences both the resonance amplitude and the linewidth. In Fig. 7 a rapid drop of the ESR amplitude diverging from the 1/T dependence is clearly visible at the temperature higher than 25 K. This means a loss of the $D^0$ electrons which according to the proposed model are thermally excited to the $D^-$ band.

The linewidth, thus electron spin relaxation, seems to be even more influenced by a presence of the $D^-$ band. The excitation to the $D^-$ states shortens electron lifetime in the $D^0$ band both when increasing temperature and increasing the doping concentration (Fig. 3). The electron lifetime in the $D^0$ band imposes an upper limit to the electron spin relaxation time $T_2$. This is a very strong effect dominating especially in samples approaching the metal-insulator transition. We will stress here, that the ESR linewidth for samples close to the MIT provides a direct measure of the electron lifetime in the $D^0$ band.

It is worth noting here, that neither from Hall effect measurements nor from the ESR we cannot unambiguously conclude about the origin of the $D^-$ band. The Mott-Hubbard model links the $D^-$ band to the upper subband originating from splitting of the donor band into single- and double-occupied states. However, one can think about other scenario, when the increased doping concentration promotes creation of native defects introducing the double-occupied states at the bottom of the GaN conduction band. In principle we cannot exclude such possibility. Though, it is established that doping with donors promotes rather deep acceptor centers to compensate the donors.[28] This effect is visible even in our electron transport analysis, where despite the increasing Si doping the compensation ratio remains roughly constant. The silicon DX centers are neither likely to be responsible for the shallow double-occupied donor band, as they have been shown to form unstable configurations.[29] So far, the splitting of the Si donor band into single- and double-occupied states gives the most plausible explanation. This terminology has been used throughout this paper.

## 7. CONCLUSIONS

A number of theoretical approaches for metal-insulator transition has been developed.[6] In this communication we have shown that the real metal-insulator transition in GaN:Si shows many features characteristic for the Mott-Hubbard type of the MIT: existence of a paramagnetic $D^0$ band of single-occupied donor sites (27 meV below the bottom of the GaN conduction band) and a non-magnetic $D^-$ band of doubly-occupied electronic states (2.7 meV below the bottom of the GaN conduction band). The metal-insulator transition corresponds to simultaneous closing the gap between the two subbands and vanishing of the activation energy for electron transport in the GaN conduction band . This occurs at the critical Si concentration equal to about $n_{Si}$-$n_a$ = 1.6 × $10^{18}$ $cm^{-3}$. When approaching the MIT, damping of the electronic magnetic moment occurs due to double occupation of electronic states. Finally, for metallic samples beyond the MIT the magnetic moment vanishes.

New spin relaxation mechanism has been observed in samples approaching the MIT. The electron spin relaxation time $T_2$ is constrained by the limited electron lifetime in the $D^0$ band, shortening both when increasing temperature and the Si concentration. It is the dominant spin relaxation mechanism at this doping level. The ESR linewidth provides in this case a direct measure of the electron lifetime in the $D^0$ band. Above the metal-insulator transition it seems that there is no use of talking about electron spin relaxation, as the conducting electrons fill double-occupied electronic states with anti-parallel coupled spins.

ACKNOWLEDGEMENTS

This work has been supported by funds for science, grant numbers: PBZ/MNiSW/07/2006/39 and N N202 1058 33, Poland.

FIGURES:

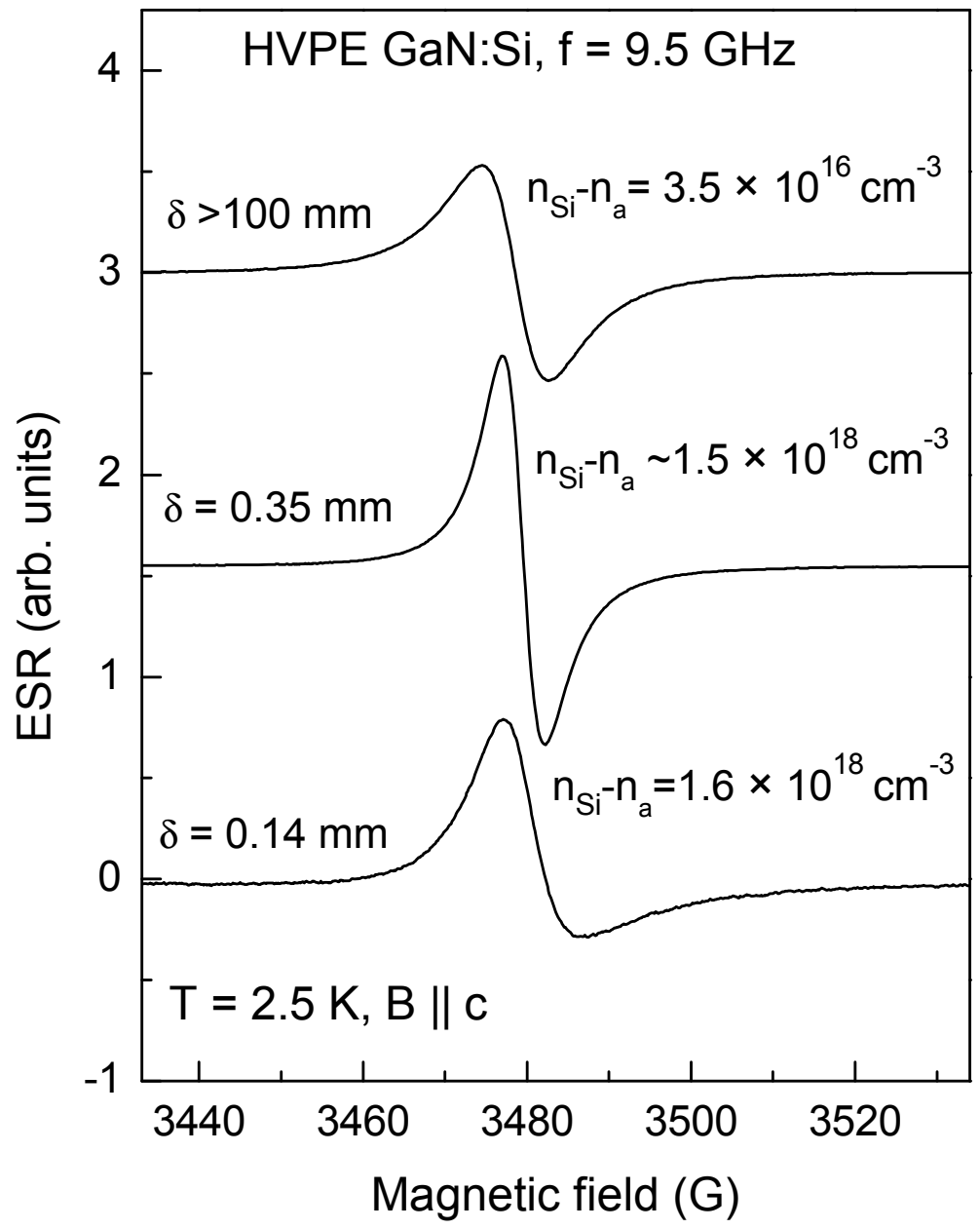

FIG. 1: ESR spectra of neutral Si donors ($D^0$) in HVPE-grown GaN. The signal evolves from Lorentzian lineshape in isolating samples to Dysonian lineshape in conducting samples. $n_{Si}$

denotes total Si concentration, $n_a$ concentration of acceptors. δ stays for the skin depth of microwave penetration.

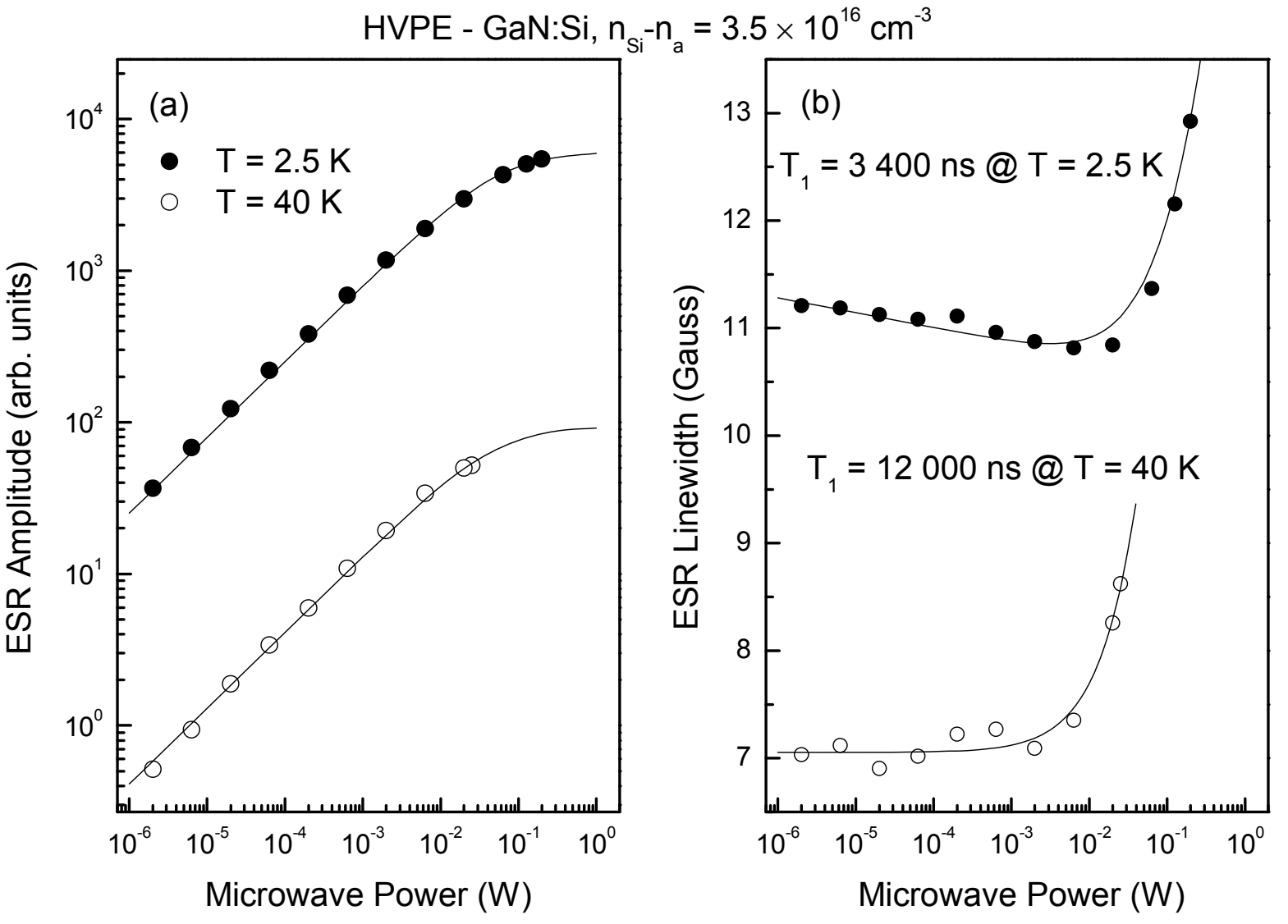

FIG. 2: Determination of the neutral donor electron spin relaxation time $T_1$ from a standard saturation method in a sample having $n_{Si}$-$n_a$ = 3.3 × $10^{16}$ $cm^{-3}$. The dependence of the amplitude (a) and the linewidth (b) of the ESR signal on microwave power is modeled with standard formulas[16,17] yielding $T_1$ equal to 3 400 ns at T = 2.5 K, and 12 000 ns at T = 40 K, respectively. The inhomogenously broadened resonance line recorded at T = 2.5 K undergoes slight narrowing with increasing microwave power below saturation. This effect is not visible in homogeneous broadened line recorded at T = 40 K

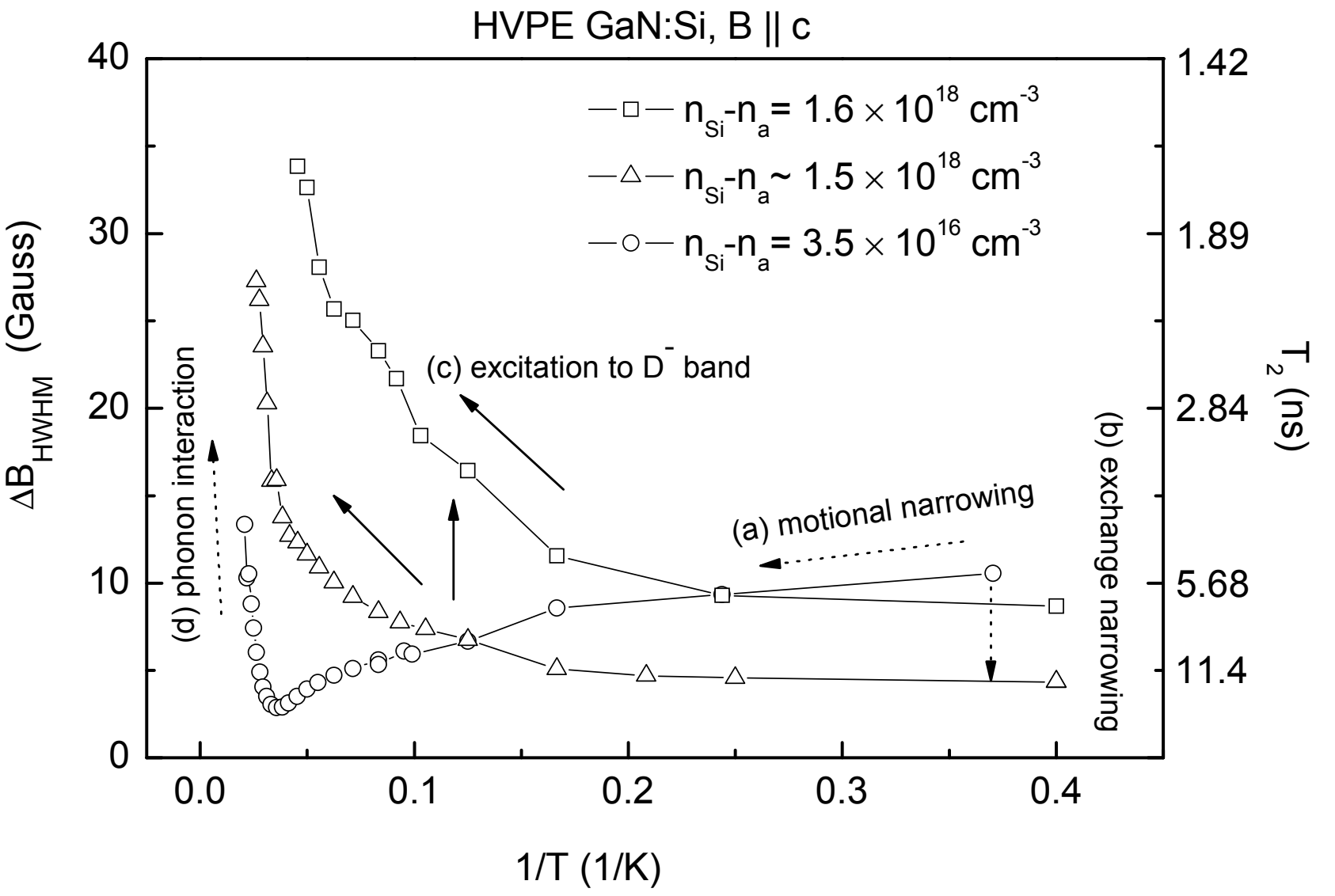

FIG. 3: Temperature dependence of the resonance linewidth (and corresponding spin relaxation time $T_2$ - right vertical scale) for HVPE-grown GaN:Si. Qualitative analysis of the dominating spin relaxation mechanisms is indicated with arrows: (a) hopping leads to the motional narrowing with increasing temperature in lightly doped samples, (b) exchange narrowing mechanism elongates electron spin relaxation time $T_2$ when increasing Si doping, (c) excitation to double occupied donor band $D^-$ shortens electron spin relaxation time both when increasing temperature and approaching critical concentration for the metal-insulator transition, (d) interaction with GaN lattice phonons shortens the $T_2$ at high temperatures.

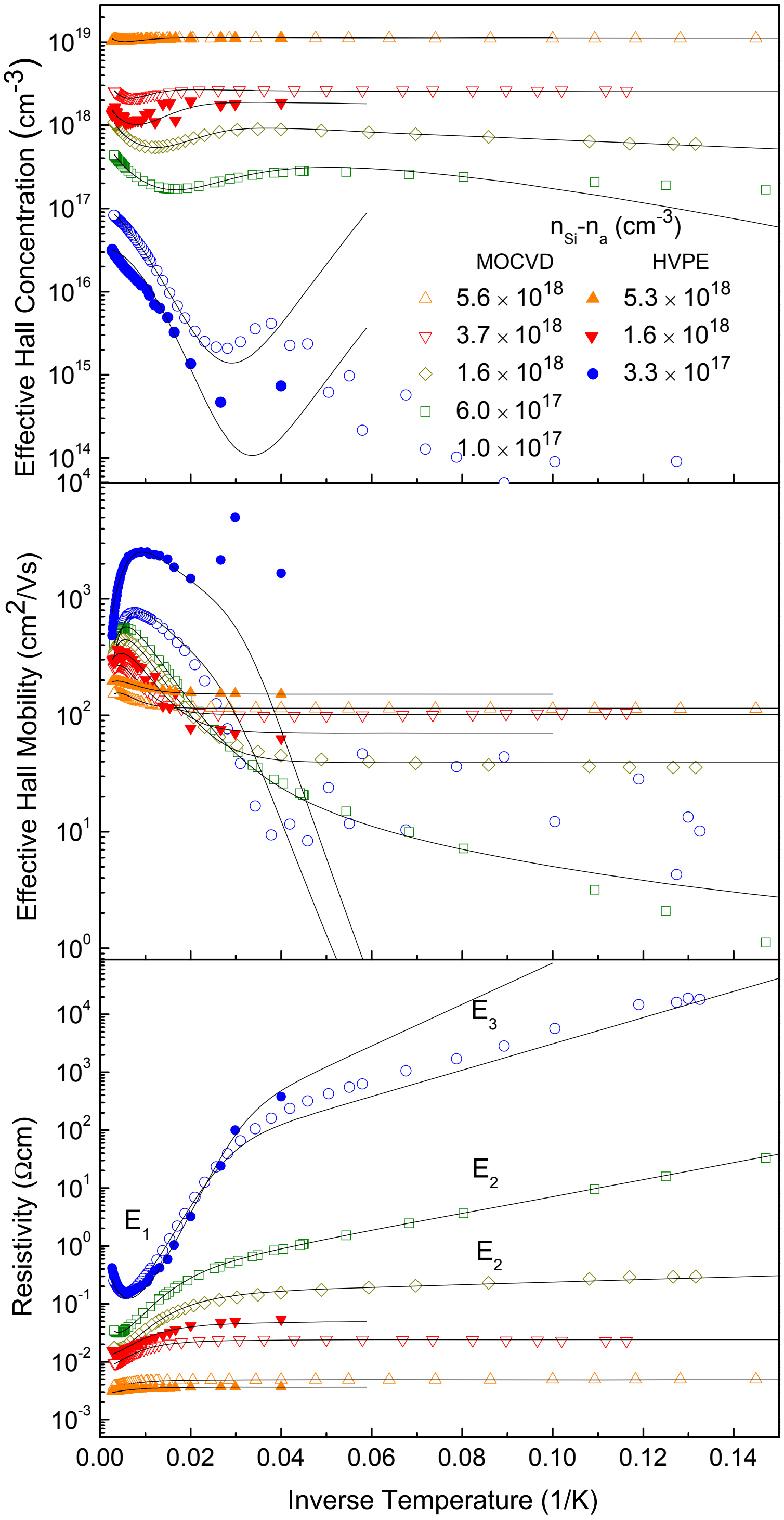

FIG. 4: (a) Effective concentration, (b) mobility, and (c) the resistivity determined by Hall measurements for a series of MOCVD- and HVPE-grown GaN:Si.

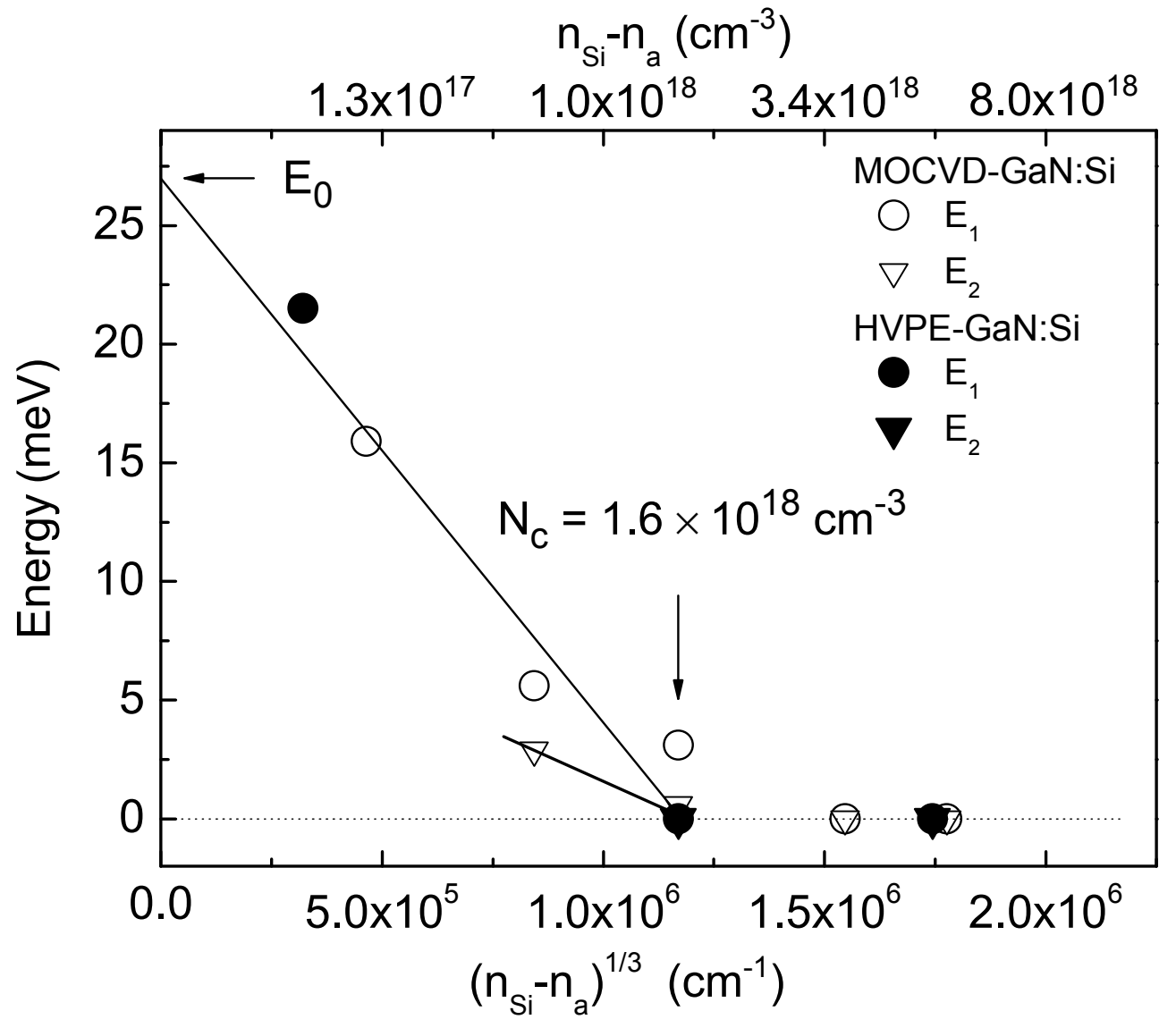

FIG. 5: Conduction band transport activation energy, $E_1$, and activation energy to double-occupied donor band, $E_2$, are plotted versus reciprocal distance between uncompensated Si donors, $n_{Si}$-$n_a$. $E_0$ denotes conduction band activation energy at low doping limit, corresponding to the location of isolated donor level with respect to the GaN conduction band. The critical concentration for the metal-insulator transition, corresponding to vanishing of both the activation energies, $E_1$ and $E_2$, equals to $N_c = 1.6 \times 10^{18}$ $cm^{-3}$. Solid lines: upper line is a fit of Eq. 1, $n_{Si}$-$n_a$ equal to $n_{donor}$ was assumed; lower line is a guide to the eye.

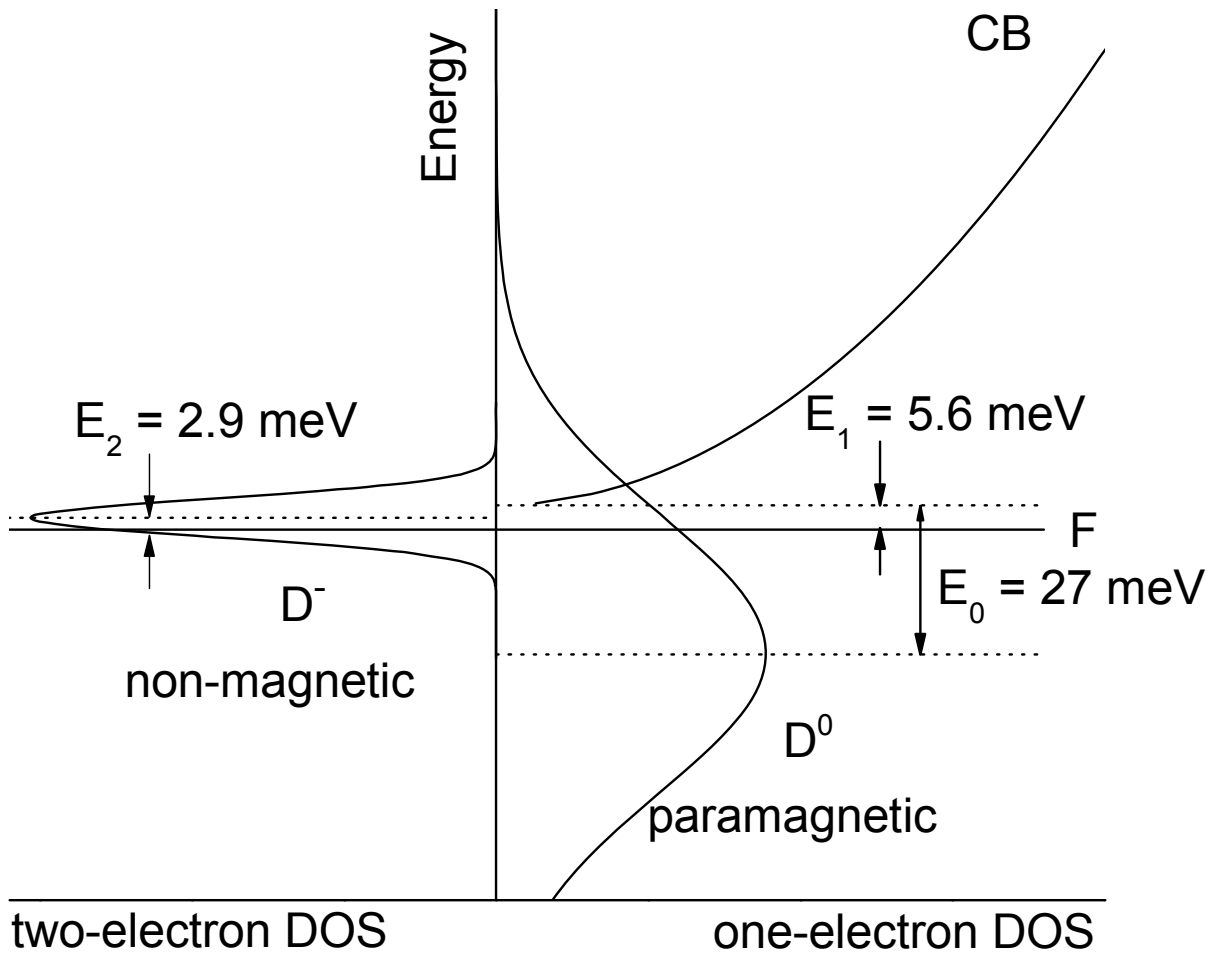

FIG. 6: Schematic diagram of the density of states (DOS) in GaN:Si with $n_{Si}$-$n_a$= $6.0 \times 10^{17}$ $cm^{-3}$. One-electron density of states are shown on the right panel, including GaN conduction band (CB) and the $D^0$ band of single-occupied Si donors. Left panel

represents two-electron density of states with $D^-$ band of double-occupied donors. F stays for the Fermi level. Conduction band activation energy, $E_1$, activation energy to double-occupied donor band, $E_2$, and activation energy of an isolated Si donor, $E_0$, are determined from Hall experiment.

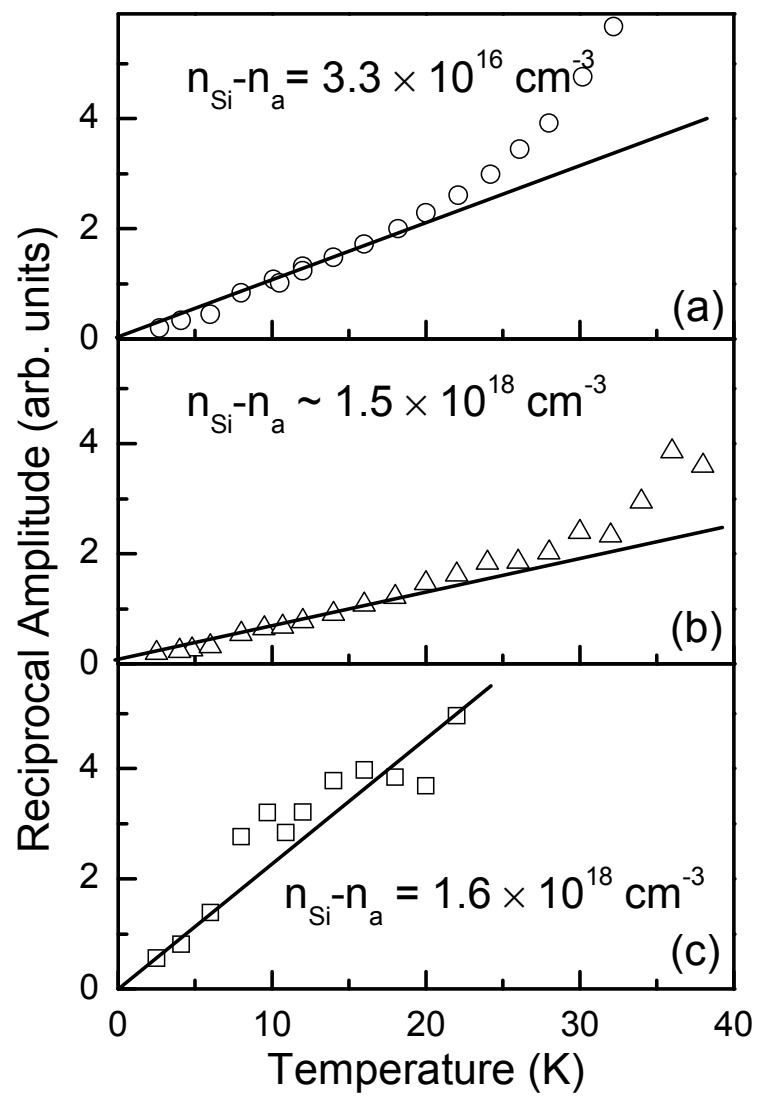

FIG. 7: Temperature dependence of ESR amplitudes for HVPE-grown GaN:Si with $n_{Si}$-$n_a$ equal to (a) 3.3 × $10^{16}$ $cm^{-3}$, (b) 1.5 × $10^{18}$ $cm^{-3}$, and (c) 1.6 × $10^{18}$ $cm^{-3}$. Solid lines are guides to the eye. Linear dependence of the reciprocal amplitude on temperature indicates Curie paramagnetism.

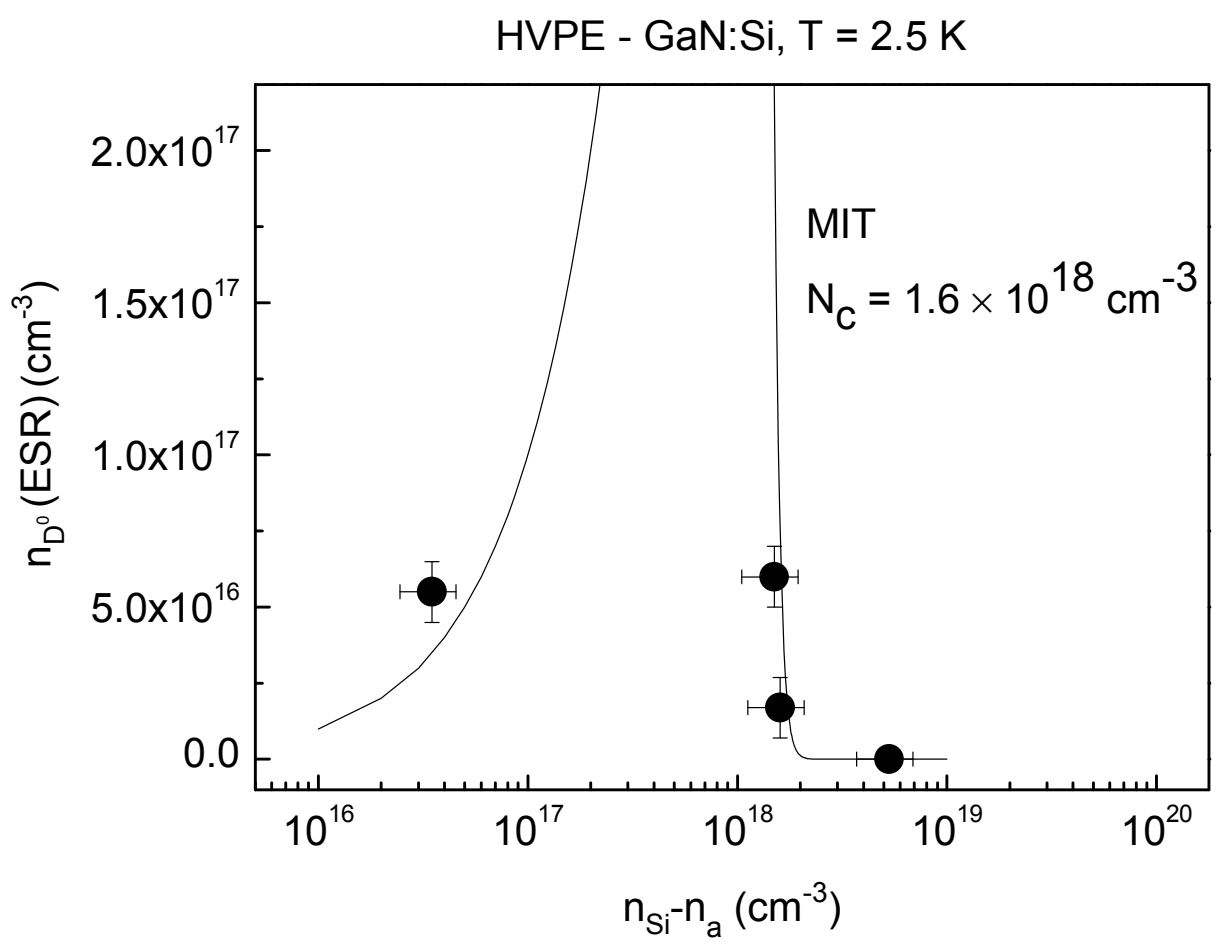

FIG. 8: Concentration of neutral Si donors determined from the amplitude of the ESR signal versus the $n_{Si}$-$n_a$. Solid line is given by the Gibbs distribution, Eq.2. At low Si concentrations the amplitude of the ESR signal scales linearly with $n_{Si}$-$n_a$. When approaching metal-insulator

transition mass occupation of $D^-$ states occurs, resulting in rapid drop of the ESR signal amplitude.

TABLE:

Table I. Electric transport parameters of MOCVD- and HVPE-grown GaN:Si determined from Hall experiment: $n_{Si}$ is Si donor concentration, $n_a$ acceptor concentration, $K = n_a/n_{Si}$ denotes compensation ratio, $E_1$, $E_2$, and $E_3$ are conduction band transport activation energy, activation energy to double-occupied donor band, $D^-$, and hopping activation energy, respectively. * denotes a sample, for which we were unable to determine transport parameters from Hall data due to parasite currents. In that case we relied on resistivity values determined from Dysonian lineshape of the ESR signal, in particular the $E_2$ was determined in this way, and on the technological data.

| $n_{Si}$-$n_a$ ($cm^{-3}$) | $K=n_a/n_{Si}$ | $E_1$ (meV) | $E_2$ (meV) | $E_3$ (meV) |
|---|---|---|---|---|
| MOCVD-GaN:Si | | | | |
| $1.0\times10^{17}$ | 0.37 | 15.9 | - | 4.5 |
| $6.0\times10^{17}$ | 0.28 | 5.6 | 2.9 | - |
| $1.6\times10^{18}$ | 0.18 | 3.1 | 0.6 | - |
| $3.7\times10^{18}$ | 0.28 | 0 | 0 | - |
| $5.6\times10^{18}$ | 0.26 | 0 | 0 | - |
| HVPE-GaN:Si | | | | |
| $3.3\times10^{16}$ | 0.30 | 21.5 | - | 7.1 |
| $\sim1.5\times10^{18}$ * | - | - | 0.6 | - |
| $1.6\times10^{18}$ | 0.27 | 0 | 0 | - |
| $5.3\times10^{18}$ | 0.46 | 0 | 0 | - |

[20] A. Wolos, M. Piersa, G. Strzelecka, K. P. Korona, A. Hruban, and M. Kaminska, Phys. Status Solidi C **6**, 2769 (2009).

[21] Yu. A. Gurvich, A. P. Melnikov, L. N. Shestakov, and E. M. Gershenzon, Laser Phys. **8**, 541 (1998).

[22] E. A. Davis and W. D. Compton, Phys. Rev. **140**, A2183 (1965).

[23] N. F. Mott, *Metal-Insulator Transitions,*(Taylor & Francis, New York, Philadelphia, 1990).

[24] Y. J. Wang, H. K. Ng, K. Doverspike, D. K. Gaskill, T. Ikedo, I. Akasaki, and H. Amano, J. Appl. Phys. **79**, 8007 (1996).

[25] W Gotz, N. M. Johnson, C. Chen, H. Liu, C. Kuo, and W. Imler, Appl. Phys. Lett. **68**, 3144 (1996).

[26] D. A. Woodbury and J. S. Blakemore, Phys. Rev. B **8**, 3803 (1973).

[27] V.L. Bonch-Bruevich and S.G. Kalashnikov, *Fizika poluprovodnikov* (Physics of Semiconductors), (Nauka, Moscow, 1977).

[28] P. Boguslawski, E. L. Briggs, and J. Bernholc, Phys. Rev. B **51**, 17255 (1995).

[29] P. Boguslawski and J. Bernholc, Phys. Rev B **56**, 9496 (1997).